\documentclass{article}
\usepackage{emulateapj}
\usepackage{psfig}

\def\ltsima{$\; \buildrel < \over \sim \;$}
\def\gtsima{$\; \buildrel > \over \sim \;$}
\def\proptosima{$\; \buildrel \propto \over \sim \;$}
\def\simlt{\lower.5ex\hbox{\ltsima}}
\def\simgt{\lower.5ex\hbox{\gtsima}}
\def\simpropto{\lower.5ex\hbox{\proptosima}}

\makeatletter
\let\@internalcite\cite
\def\cite{\def\astroncite##1##2{##1\ ##2}\@internalcite}
\def\citey{\def\astroncite##1##2{##1\ (##2)}\@internalcite}
\def\@citex[#1]#2{\if@filesw\immediate\write\@auxout{\string\citation{#2}}\fi
  \def\@citea{}\@cite{\@for\@citeb:=#2\do
    {\@citea\def\@citea{; }\@ifundefined
       {b@\@citeb}{{\bf ??}\@warning
       {Citation `\@citeb' on page \thepage \space undefined}}%
{\csname b@\@citeb\endcsname}}}{#1}}
\def\@cite#1#2{#1\if@tempswa #2\fi}
\def\@biblabel#1{}
\def\astroncite#1#2{#1\ #2}
\makeatother

\begin{document}

\newcommand{\etal}{et al.}

\slugcomment{To appear in April 10, 2000 ApJ Part 1}

\lefthead{Y.Tsuboi et al.}
\righthead{YLW15}


\title{Quasi-periodic X-ray Flares from the Protostar YLW15}

\author{Yohko Tsuboi\altaffilmark{1}, Kensuke Imanishi and Katsuji Koyama\altaffilmark{2}}
\affil{Department of Physics, Faculty of Science, Kyoto University,
Sakyo-ku, Kyoto 606-8502, Japan}
\affil{tsuboi@cr.scphys.kyoto-u.ac.jp, kensuke@cr.scphys.kyoto-u.ac.jp,
koyama@cr.scphys.kyoto-u.ac.jp}

\and

\author{Nicolas Grosso and Thierry Montmerle}
\affil{Service d'Astrophysique, CEA/DAPNIA/SAp, Centre d'Etudes de Saclay,
91191 Gif-sur-Yvette Cedex, France}
\affil{grosso@discovery.saclay.cea.fr, mtmerle@discovery.saclay.cea.fr}

\altaffiltext{1}{Present address: Department of Astronomy \& Astrophysics, The 
Pennsylvania State University, University Park, PA 16802-6305, USA}
\altaffiltext{2}{CREST, Japan Science and Technology Corporation
(JST), 4-1-8 Honmachi, Kawaguchi, Saitama, 332-0012, Japan}

\begin{abstract}

With {\it ASCA}, we have detected three X-ray flares from the Class I
protostar YLW15. The flares occurred every $\sim$20 hours and showed an
exponential decay with time constant 30--60 ks. The X-ray spectra are
explained by a thin thermal plasma emission. The plasma temperature
shows a fast-rise and slow-decay for each flare with $kT_{peak} \sim$
4--6 keV. The emission measure of the plasma shows this time profile
only for the first flare, and remains almost constant during the
second and third flares at the level of the tail of the first flare.
The peak flare luminosities $L_{X,peak}$ were $\sim$ 5--20 $\times
10^{31}$ erg s$^{-1}$, which are among the brightest X-ray
luminosities observed to date for Class I protostars. The total energy
released in each flare was 3--6$\times$10$^{36}$ ergs. The first flare
is well reproduced by the quasi-static cooling model, which is based
on solar flares, and it suggests that the plasma cools mainly
radiatively, confined by a semi-circular magnetic loop of length $\sim
14\,R_\odot$ with diameter-to-length ratio $\sim 0.07$. The two
subsequent flares were consistent with the reheating of the same
magnetic structure as of the first flare. The large-scale magnetic
structure and the periodicity of the flares imply that the reheating
events of the same magnetic loop originate in an interaction between
the star and the disk due to the differential rotation.

\end{abstract}


\section{Introduction}

Low-mass Young Stellar Objects (YSOs) evolve from molecular cloud
cores through the protostellar (ages $\sim$10$^{4-5}$ yr), the
Classical T Tauri (CTTS: $\sim$ 10$^6$ yr) and Weak-lined T Tauri
(WTTS: $\sim$10$^7$ yr) phases to main sequence. Protostars are
generally associated with the Class 0 and I spectral energy
distributions (SEDs), which peak respectively in the millimeter and infrared
(IR) bands. Bipolar flows are accompanied with this phase, suggesting
dynamic gas accretion. CTTSs have still circumstellar disks though
they have expelled or accreted the infalling envelopes. They are
associated with the Class II spectra, which peak at the
near-IR. Finally, as the circumstellar disk disappears, YSOs
evolve to WTTSs, associated with Class III stars. 
Early stellar evolution is reviewed by Shu, Adams, \& Lizano
(1987) and Andr\'e \& Montmerle (1994).

The {\it Einstein} Observatory discovered that T Tauri Stars (TTSs),
or Class II and Class III infrared objects, are strong X-ray emitters,
with the luminosities of 100--10000 times larger than solar
flares. These X-rays showed high amplitude time variability like solar
flares. The temperature ($\sim$1 keV) and plasma density
($n_e\sim$10$^{11}$cm$^{-3}$) are comparable to those of the Sun,
hence the X-ray emission mechanism has been thought to be a scaled-up
version of solar X-ray emission; i.e., magnetic activity on the
stellar surface enhanced by a dynamo process (Feigelson \& DeCampli
1981; Montmerle et al. 1983). X-ray and other high energy processes in
YSOs are reviewed by Feigelson \& Montmerle (1999).

In contrast to TTSs, Class I infrared objects are generally surrounded
by circumstellar envelopes of $A_V$ up to $\sim$ 40 or more, hence are
almost invisible in the optical, near infrared and even soft X-ray
bands. The {\it ASCA} satellite, sensitive to high energy X-rays up to
10 keV which can penetrate heavy absorption, has found X-rays from
Class I objects in the cores of the R CrA, $\rho$ Oph, and Orion
clouds at the hard band ($>$ 2 keV) (Koyama et al. 1994; Koyama et
al. 1996; Kamata et al. 1997, Ozawa et al. 1999). Even in the soft
X-ray band, deep exposures with the {\sl ROSAT} Observatory detected
X-rays from YLW15 in $\rho$ Oph (Grosso et al. 1997) and CrA
(Neuh\"auser \& Preibisch 1997).

A notable aspect of these pioneering observations was the discovery of
X-ray flares from Class I stars. {\sl ROSAT} discovered a giant flare
from the protostar YLW15 with total luminosity (over the full X-ray
band) of 10$^{34-36}$erg s$^{-1}$, depending on the absorption. {\sl
ASCA} observed more details of X-ray flares from protostars EL29 in the
Opiuchus, R1 in the R CrA core, and SSV63E+W in the Orion, which are
associated with larger $N_H$ $\simeq$ 10$^{22-23}$ cm$^{-2}$ than seen
in TTSs.

All these findings led us to deduce that greatly enhanced magnetic
activity, already well-established in older TTSs, is present in the
earlier protostellar phase. 
Stimulated by these results, and to search for further examples of the
protostellar activity in the X-ray band, we have performed an
extensive follow-up observation of a core region in $\rho$ Oph, with
several Class I X-ray sources. The follow-up observation was made with
{\it ASCA} 3.5 years after the first observation (Koyama et al. 1994,
Kamata et al. 1997). Some previously bright Class Is became dim, while
other Class Is were identified as hard X-ray sources. This paper
discusses the brightest hard X-ray source, YLW15, concentrating on the
characteristics and implications of its peculiar time behavior:
quasi-periodic hard X-ray flares. For comparison with previous
results, we assume the distance to the $\rho$ Oph region
to be 165 pc (Dame et al. 1987) throughout of this paper,
although new Hipparcos data suggest a closer distance $d\sim120$ pc
(Knude \& Hog 1998).

\section{Observation}

We observed the central region of $\rho$ Oph cloud with {\it ASCA} for
$\approx$100 ks on 1997 March 2--3. The telescope pointing coordinates
were $\alpha$(2000) = 16h 27.4m and $\delta$(2000) = $-$24${^\circ}$
30$'$. All four detectors, the two Solid-state Imaging Spectrometers
(SIS 0, SIS 1) and the two Gas Imaging Spectrometers (GIS 2, GIS 3)
were operating in parallel, providing four independent data
sets. Details of the instruments, telescope and detectors are given by
Burke et al. (1991), Tanaka, Inoue, \& Holt (1994), Serlemitsos et
al. (1995), Ohashi et al. (1996), Makishima et al. (1996), and
Gotthelf (1996).

Each of the GIS was operated in the Pulse Height mode with the standard
bit assignment that provides time resolutions of 62.5 ms and 0.5 s for
high and medium bit rates, respectively. The data were post-processed
to correct for the spatial gain non linearity. Data taken at
geomagnetic rigidities lower than 6 GV, at elevation angles less than
5$^\circ$ from the Earth, and during passage through the South
Atlantic Anomaly were rejected. After applying these filters, the net
observing time for both GIS 2 and GIS 3 was 94 ks.

Each of the SIS was operated in the 4-CCD/Faint mode (high bit rate)
and in the 2-CCD/Faint mode (medium bit rate). However, we concentrate
on the 4-CCD/Faint results in this paper since YLW15 is out of the
2-CCD field of view. The data were corrected for spatial and gain
non-linearity, residual dark distribution, dark frame error, and hot
and flickering CCD pixels using standard procedures.
Data were rejected during South Atlantic
Anomalies and low elevation angles as with GIS data.
In order to avoid contamination due to
light leaks through the optical blocking filters, we excluded data
taken when the satellite viewing direction was within 20$^\circ$ of
the bright rim of the Earth. After applying these filters, the net observing
time for the 4-CCD mode was 61 ks for SIS 0 and 63 ks for SIS 1.

Towards the end of this observation, we detected an enormous flare
from T Tauri star ROXs31 which is located close to YLW15 (see \S3.1,
source 6 in Figure 1). The peak flux of ROXs31 is 1--2 orders of
magnitude larger than YLW15 (see \S4.2), and 
its broad point spread function contaminates YLW15 
during the flare. Therefore, we excluded the GIS and SIS data taken during the
flare of ROXs31 in all analysis of YLW15.

\section{Results and Analysis}

\subsection{Images}

Figure 1 shows X-ray images of the $\rho$ Ophiuchi Core F region in the two
different energy bands (left panel: 0.7--2 keV, right panel: 2--10
keV), obtained with SIS detectors. Class I sources are indicated by
crosses. Since the absolute {\it ASCA} positional errors can be as
large as 40$''$ for SISs (Gotthelf 1996), we compared the {\it ASCA}
peak positions to the more accurately known IR positions of two bright
sources in the SIS field, ROXs21 (source 5)
and ROXs31 (source 6), which are indicated by filled circles in Figure
1 left panel. To obtain the {\it ASCA} peak positions, we executed
a two-dimensional fitting in the 0.7--2 keV band; we fitted these sources
with a position-dependent point spread function in the 0.7--2 keV band
and a background model. This procedure was done in the Display45 analysis
software package (Ishisaki et al. 1998).
The position
of ROXs31 was based on the flare phase of this source, while the
position of ROXs21 is based on the data before the flare of
ROXs31. IR positions are provided by Barsony et
al. (1997). The {\it ASCA} SIS positions had an average offset
(weighted mean by photon counts) of $+$0.18 s in right ascension and
$-$7.4$''$ in declination from the IR frame. This positional
offset is corrected in Figure 1. After the boresight error correction,
remaining excursions between the X-ray and IR positions are 5.5$''$
(rms), which is consistent with the SIS position uncertainty for point
sources (Gotthelf 1996). We take the systematic positional
error to be 5.5$''$.

From the 2--10 keV band image, we find that the X-ray fluxes from
Class I protostars EL29 and WL6 (sources 3 and 4 in Figure 1, right
panel) are fainter by one third and less than one third, respectively,
comparing to those in the first {\it ASCA} observation made in August 1993
(Kamata et al. 1997). The brightest X-ray source in the 2--10 keV band
is an unresolved source at $\alpha$(2000) = 16h 27m 27.0s and
$\delta$(2000) = $-$24$^\circ$ 40$'$ 50$''$ in the position corrected
frame.  We derived this peak position by the two-dimensional fitting
in 2--10 keV band.  Since the statistical error is 1$''$, the overall
X-ray error (including the systematic error) is $\pm6''$.

The closest IR source is YLW15 with VLA position of $\alpha$(2000) =
16h 27m 26.9s and $\delta$(2000) = $-$24$^\circ$ 40$'$ 49.8$''$
($\pm0.5''$; Leous et al. 1991), located 1.5$''$
away from the X-ray source. The next nearest source is GY263 with IR
position of $\alpha$(2000) = 16h 27m 26.6s and $\delta$(2000) = $-$24$^\circ$
40$'$ 44.9$''$ ($\pm$1.3$''$; Barsony et
al. 1997). The source is located 5.5$''$ from the X-ray position
on the border of the X-ray position error circle. Thus we
conclude that the hard X-rays are most likely due to the Class I
source YLW15.


\vspace{0.25cm}
\begin{figure*}
\centerline{\hbox{\psfig{file=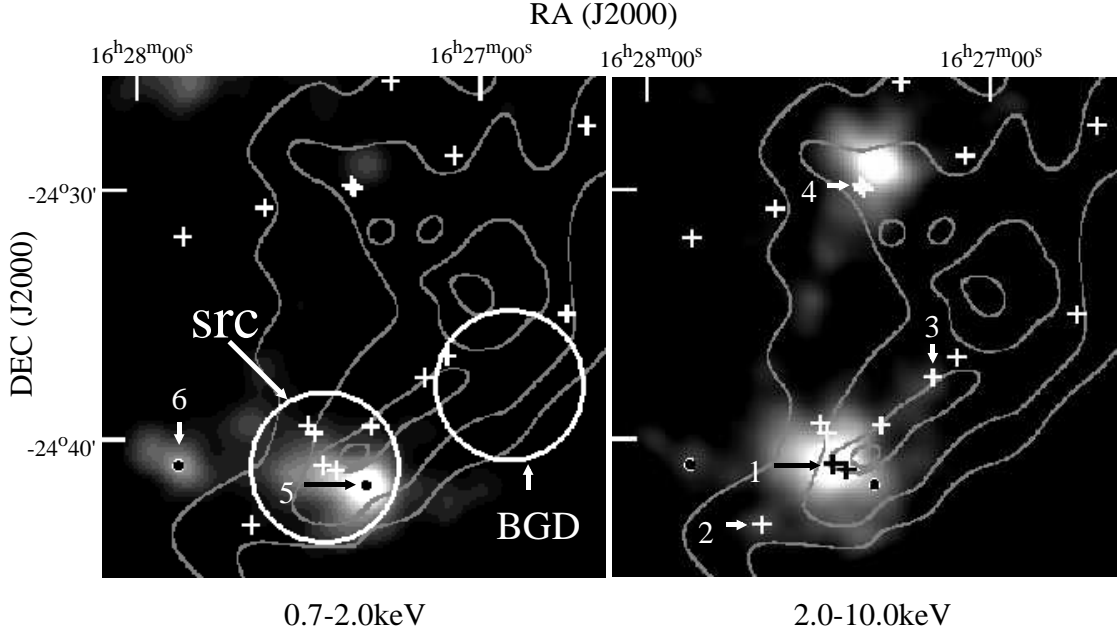,width=0.8\textwidth}}}
\figcaption{\small 
The SIS images of the central region of $\rho$ Oph cloud
in the 0.7--2 keV (left) and the 2--10 keV band (right).
The contours are C$^{18}$O(1$\rightarrow$0) column densities
(Wilking \& Lada 1983)
in units of 10$^{16}$cm$^{-2}$, from 1.0 to 2.5 in steps of 0.5,
corresponding to $A_V\sim$30, 50, 70, 100, respectively.
Class I sources are indicated by crosses (Chen {\etal} 1995, 1997;
Motte, Andr\'e, \& Neri 1998). The bright sources 5 and 6 in the
0.7--2 keV band are T Tauri star ROXs21 and ROXs31, respectively. In
the 2--10 keV band image, the brightest source is Class I source YLW15
(source 1). Two older Class I sources, IRS51 (source 2) and EL29
(source 3), which have flat SEDs are also detected with a criterion of
S/N $\geq$ 5$\sigma$. EL29 and WL6 (source 4) are faint, though
both of them emitted strong X-rays in the first {\it ASCA} observation
(Koyama et al. 1994, Kamata et al. 1997). The hard X-ray source near
WL6 is a cluster of T Tauri stars (WL3, WL4, WL5; Wilking \&
Lada 1983). The source region and background region for YLW15 are also
shown by circles.}
\end{figure*}
\vspace{0.25cm}

\subsection{X-ray Lightcurve of YLW15}

We extracted a lightcurve from a $3'$ radius circle around the X-ray
peak of YLW15 (see Figure 1). Before the enormous flare from T Tauri
star ROXs31, which occurred in the last phase of this observation (see
\S 2), we detected another large flare from Class II source SR24N,
located about $7'$ away from YLW15 in the GIS field of view. To
subtract the time variable contamination from the SR24N in the
extended flux of YLW15, we selected a $3'$ radius background region
(see Figure 1), equidistant from SR24N and YLW15. On the other hand,
using such a background, we cannot exclude the contamination from
ROXs21 (source 5 in Figure 1), which is 2 arcmin apart from
YLW15. Since the X-rays from ROXs21 are dominant below 2 keV (see \S 3.3
and Fig.3), we 
examine time variability only in the hard X-ray band ($>$ 2 keV)
in which the flux is dominated by YLW15.

Figure 2 (upper panel) shows the background-subtracted lightcurve in
the 2--10 keV band with the sum of the SIS (SIS 0 and 1) and GIS (GIS
2 and 3) images. The lightcurve shows a sawtooth pattern with three
flares. The peak fluxes of the flares become successively less
luminous. Each flare exhibits a fast-rise and an exponential decay
with an $e$-folding time of 31$\pm1$ ks ($\chi^2/d.o.f.$ = 61/46),
33$\pm3$ ks (80/47), and 58$_{-13}^{+24}$ ks (33/24), for the first,
the second, and the third flares, respectively. We show the best fit
lightcurves for the second and the third flares with dashed lines, and
show the best-fit quasi-static model (see \S 4.1.1) for the first
flare with a solid line in Figure 2 upper panel.


\vspace{0.25cm}
\centerline{\hbox{\psfig{file=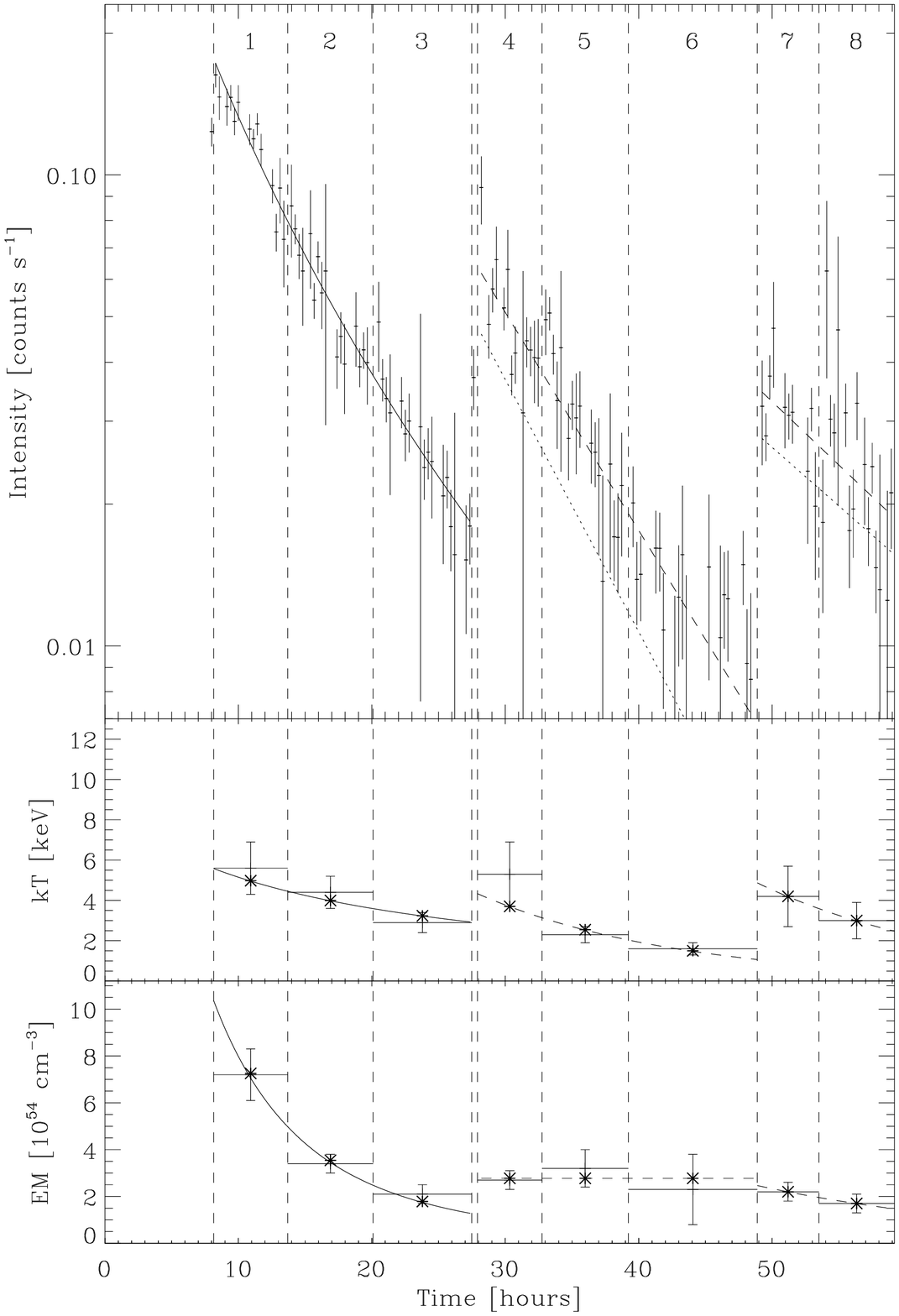,width=0.4\textwidth}}}
\figcaption{\small
Upper panel: Background-subtracted lightcurve of
YLW15 obtained with all the detectors (SIS 0 + SIS 1 + GIS 2 + GIS 3)
in the 2--10 keV band. All time bins are 1024 s wide.  Middle panel:
Time profile of the best-fit temperature. The error bars indicate 90\%
confidence.  Lower panel: Same as the middle panel, but for the
best-fit emission measure.  The solid lines during Flare I are the
best-fit quasi-static model, and the dashed lines during Flare II and III
are the best-fit exponential model (see Table 2). One reheating plasma
loop is assumed in both cases. Dotted lines during Flare II and III in
the upper panel show each flare component without the contribution of
the other flares, under assumption that the three flares occurred
independently (see \S4.1). Asterisks show the average of the model during
one time bin.  The time axis starts at 00:00:00.0 UT (1997 March 2).
}
\vspace{0.25cm}


\subsection{Time-Sliced Spectra of YLW15}

To investigate the origin of the quasi-periodic flares, we made
time-sliced spectra for the time intervals given in Figure 2. We
extracted the source and background data for each phase from the same
regions as those in the lightcurve analysis (see \S 3.2 and Figure 1).
We found that all the spectra show a local flux
minimum at $\approx$1.2 keV. For example, we show the
spectra obtained with SISs at phases 1 and 8
in Figure 3.  This suggests that the spectra have two
components, one hot and heavily absorbed, the other cool and less
absorbed.

Then we examined possible contamination from the bright, soft X-ray
source ROXs21 (source 5 in Figure 1). We extracted the spectrum of ROXs21
from a $2'$ radius circle around its X-ray peak. We extracted the data
only during phase 1, in order to be free from contamination from the
flare on SR24N, which occurred during phases 4--6 (see \S 3.2). The
background data for ROXs21 were extracted from a $2'$ radius circle
equidistant from YLW15 and ROXs21 during phase 1. After the
subtraction of the background, the spectrum of ROXs21 is
well reproduced by an optically thin thermal plasma model of about 0.6
keV temperature with absorption fixed at $N_{\rm H} =
$1.3$\times$10$^{21}$ cm$^{-2}$ ($A_V$ = 0.6 mag; Bouvier and
Appenzeller 1992) with $\chi^2/d.o.f.$ = 17/23. The flux of the soft
component of YLW15 is about 30\% of the flux of ROXs21, which is equal
to the spill-over flux from ROXs21. Thus the soft X-ray component
found in YLW15 spectra is due to contamination from the nearby bright
source ROXs21.

Having obtained the best-fit spectrum for ROXs21, we fitted the
spectrum of YLW15 in each phase with a two-temperature thermal plasma
model. The cool component is set to the
contamination from ROXs21, and the hot component is from YLW15. For
YLW15, free parameters are temperature ($kT$), emission measure ($EM$),
absorption ($N_{\rm H}$) and metal abundance. For ROXs21, 
$EM$ is the only free
parameter and the other parameters are fixed
to the best-fit values obtained in phase 1. We found no significant
variation in $N_{\rm H}$ of YLW 15 from phase to phase, hence, we fixed
the $N_{\rm H}$ to the best-fit value at phase 1. The resulting
best-fit parameters of YLW15 for each time interval are shown in Table
1. The best-fit spectra of phases 1 and 8 are illustrated in Figure 3,
and the time evolution of the best-fit parameters are shown in Figure
2.



\vspace{0.25cm}
\centerline{\hbox{\psfig{file=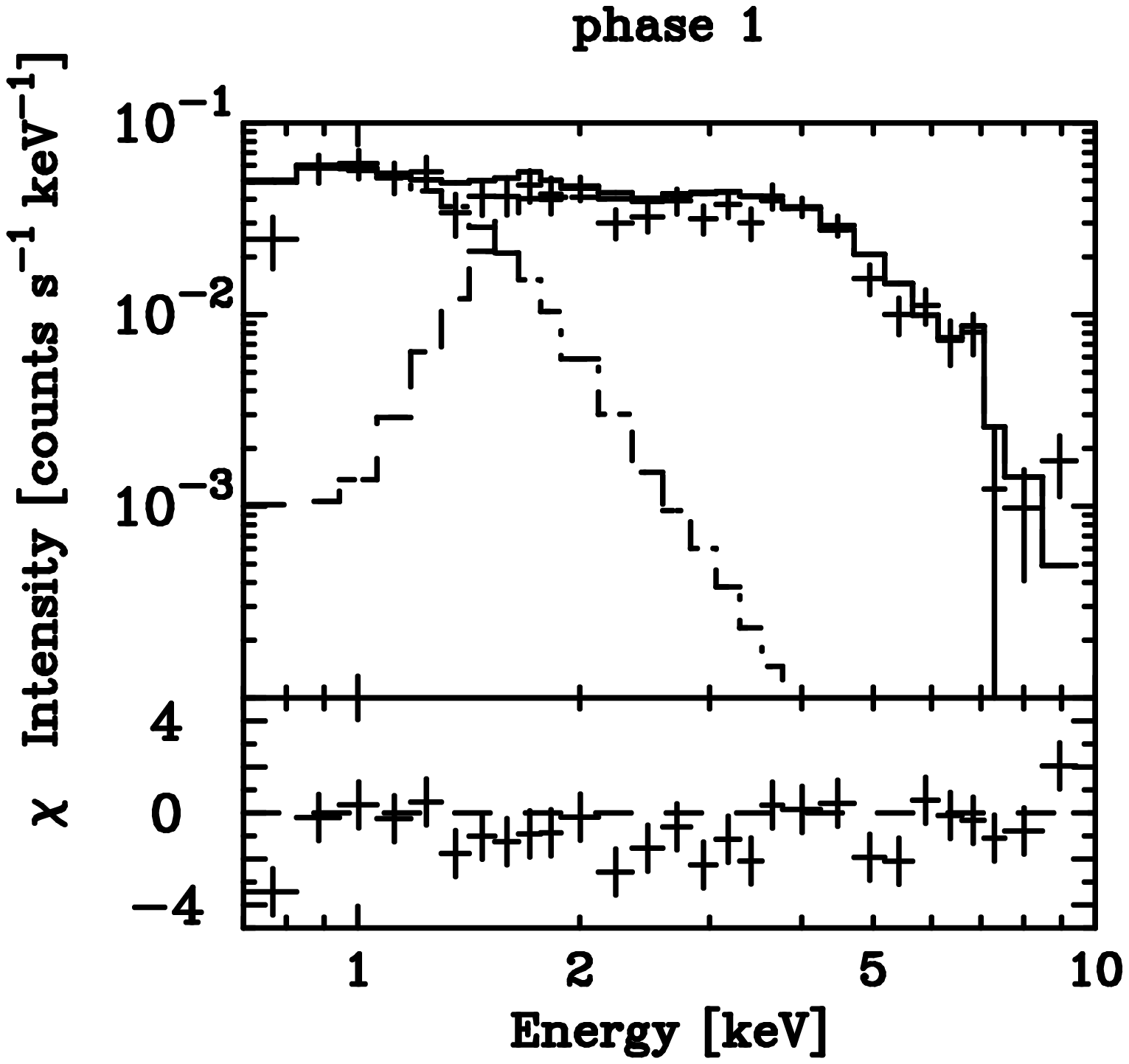,width=0.4\textwidth}}}
\centerline{\hbox{\psfig{file=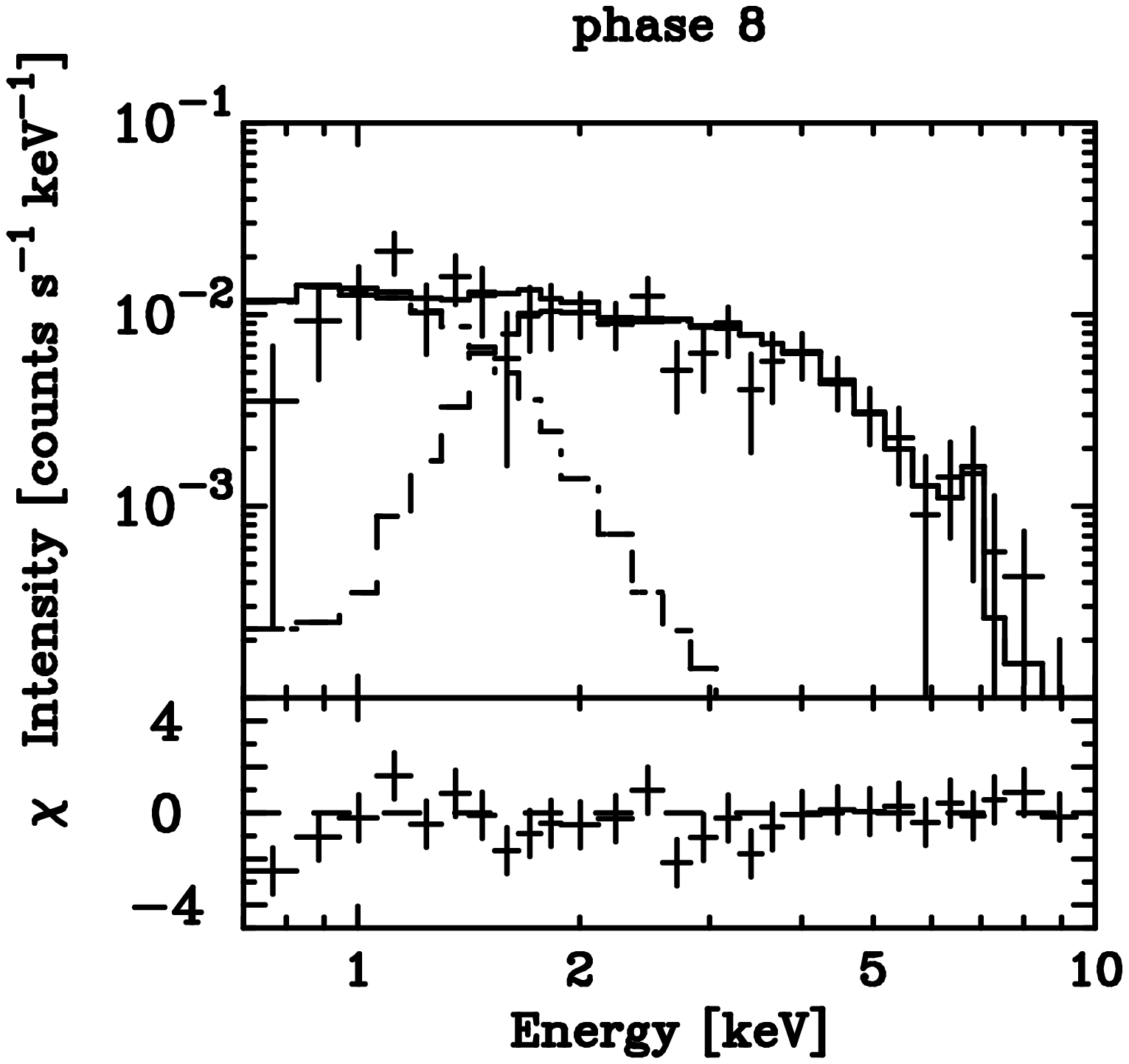,width=0.4\textwidth}}}
\figcaption{\small
The SIS (SIS 0 + 1) spectra of YLW15
at phase 1 and 8 (see Figure 2).
The solid lines show the best-fit two-temperature coronal plasma model
derived from simultaneous fitting of GIS and SIS data.
Lower panel shows the residuals from the best-fit model.
}


\small
\begin{deluxetable}{cccccccc}
\tablecaption{Best-fit parameters to the spectra of YLW15$^a$.}
\tablehead{
\colhead{Phase$^b$} & \colhead{$kT$}   & \colhead{Abundance}     & 
\colhead{$EM^c$}
  &  \colhead{$L_{\rm X}^d$}        & \colhead{$\chi^2$/d.o.f}            \\
        & \colhead{(keV)}             & \colhead{(solar)}
        & \colhead{(10$^{54}$cm$^{-3}$)}     & \colhead{(10$^{31}$erg s$^{-1}$)}    &}
\startdata
 1  & 5.6$^{+1.3}_{-1.0}$     & 0.4$^{+0.2}_{-0.1}$       & 7.2$_{-0.8 }^{+1.1}$  & 15$\pm$1   & 181/106  \\
 2  & 4.4$^{+0.8}_{-0.7}$     & 0.4$^{+0.2}_{-0.2}$ & 3.4$_{-0.3}^{+0.4}$     & 6.0$\pm$0.7    & 191/164 \\
 3  & 2.9$^{+0.5}_{-0.3}$     & 0.7$^{+0.5}_{-0.4}$ & 2.1$_{-0.3}^{+0.4}$      & 3.4$\pm$0.8    & 80/84   \\
 4  & 5.3$^{+1.6}_{-1.1}$     & 0.6$^{+0.4}_{-0.4}$  & 2.7$_{-0.3}^{+0.4}$    & 5.6$\pm$0.8   &  55/52\\
 5  & 2.3$^{+0.4}_{-0.3}$     & 0.5$^{+0.6}_{-0.4}$  & 3.2$_{-0.7}^{+0.8}$   & 4.4$\pm$1.3    & 163/138 \\
 6  & 1.6$^{+0.3}_{-0.3}$     & $<$1.3      & 2.3$_{-1.2}^{+1.5}$            & 2.5$\pm$1.5   & 46/40 \\
 7  & 4.2$^{+1.5}_{-0.9}$     & $<$0.4      & 2.2$_{-0.3}^{+0.4}$            & 3.5$\pm$0.5   & 153/139 \\
 8  & 3.0$^{+0.9}_{-0.5}$     & 0.7$^{+1.3}_{-0.5}$    & 1.7$_{-0.4}^{+0.4}$ & 2.9$\pm$0.6   & 59/65 \\
\enddata
\tablecomments{The errors are at 90\% confidence
level. \\
\hspace*{10pt}$^a$ Absorption column density is
fixed to the best fit value for phase 1, 3.3$\times10^{22}$ cm$^{-2}$. \\
\hspace*{10pt}$^b$ see Figure 2. \\
\hspace*{10pt}$^c$ Emission Measure, $\int n_e^2 dV$ 
($n_e$: electron density, $V$: emitting volume).\\
\hspace*{10pt}$^d$ Absorption-corrected X-ray luminosity in the 0.1 -- 100 keV 
band.}
\label{best_fit_parameters}
\end{deluxetable}
 
\normalsize

\section{Discussion}

In this observation, we detected hard X-rays ($>$ 2 keV) from the
Class I source YLW15 for the first time; the source was not detected in
the first {\it ASCA} observation executed 3.5 years later (Kamata et
al. 1997). On the other hand, the Class I sources EL29 and WL6, which
emitted bright hard X-rays in the first observation, became very faint
(see \S 3.1). Then we conclude that hard X-rays from Class I
protostars in the $\rho$ Oph cloud are highly variable in the time
spans, and we suspect that the non-detection of hard X-rays from other
Class I objects is partly due to the long-term time variability.
From YLW15, we discovered a peculiar time behavior: quasi-periodic
flares. We shall now discuss the relation between each flare from
YLW15 and the physical conditions.

\subsection{Physical Parameters of the Triple Flare} 

If the three intermittent flares are attributed to a single persistent
flare, with the three events due to geometrical modulation, such as
occultation of the flaring region by stellar rotation or orbit in an
inner disk, then only the emission measure should have shown three
peaks. The temperature would have smoothly decreased during the three
flares. However, in our case, we see the temperature following the
same pattern as the luminosity, decreasing after each jump, as
indicated in Table 1 and Figure 2 middle panel. To test further, we
fitted the temperatures in phases 1--8 with a single exponential decay
model, and confirmed that it is rejected with 89 \% confidence. We
then interpret the variability of YLW15 as due to a triple flare, in
which each flare followed by a cooling phase. Now, let us label phases
1--3, 4--6, and 7--8 as Flare I, II, and III, respectively. In this
subsection we will use the cooling phases to estimate the physical
conditions of the plasma in each flare. Details are given in the
Appendix.

\subsubsection{Flare I}
Here we assume that the plasma
responsible for Flare I is confined in one semi-circular magnetic loop
with constant cross section along the tube flux, with length $L$, and
diameter-to-length (aspect) ratio $a$, based on the general analogy of
solar-type flares.
To the decay of Flare I, we apply a quasi-static cooling model (van
den Oord \& Mewe 1989), in which the hot plasma cools quasi-statically
as a result of radiative (90 \%) and conductive (10 \%) losses (see
detailed comment in Appendix \S1).
If the cooling is truly
quasi-static, $T^{13/4}/EM$ shows constant during the
decay (where $T$ is plasma temperature and $EM$ is the emission
measure). We find for the three successive bins (phases 1--3) of Flare
I that $(T/10^7\,{\rm K})^{13/4}/(EM/10^{54}\,{\rm cm}^{-3}) =
61\pm55, ~59\pm42, ~25\pm18$, 
which are not inconsistent with a constant value, taking into account
the large error bars. Then, fitting our data with the quasi-static
model, we find satisfactory values of $\chi^2_{red}$ for the count
rates, temperature, and emission measure (see panel~1 in
Table~\ref{fit}). We conclude that our hypothesis of quasi-static
cooling is well verified; an underlying quiescent coronal emission is
not required to obtain a good fit. The best fits are shown by solid
lines in Figure 2. From the peak values of $T$ and $EM$, and the
quasi-static radiative time scale, we derived loop parameters as
listed in Table~\ref{physical_parameters}. The detailed procedure is
written in Appendix. The aspect ratio $a$ of 0.07 is within the
range for solar active-region loops ($0.06 \le a_\odot \le 0.2$, Golub
et al. 1980), which supports the assumed solar analogy.

\subsubsection{Flares II and III} 
The appearance of Flare II can be interpreted either by the reheating
of the plasma cooled during Flare I, or by the heating of a distinct
plasma volume, independent from that of Flare I. In the latter case,
the lightcurve and $EM$ would be the sum of Flare I component and new
flare component. The $EM$ of the new flare component can be derived by
subtracting the component extrapolated from Flare I. The derived $EM$
are 1.6$\pm$0.4 (phase 4), 2.5$\pm$0.8 (phase 5), and 1.9$\pm$1.5
(phase 6). Then we fitted the lightcurve and $EM$ with the
quasi-static model assuming the above two possibilities. However, in both of
the cases, the model did not reproduce the lightcurve and $EM$
simultaneously, and the quasi-static cooling model cannot be adopted
throughout the triple flare but Flare I.

For simplicity, we fitted the parameters in Flare II with an
exponential model. The best-fit parameters are shown in Table~2, and
each models for the reheating and the distinct flare assumptions are
shown by the dashed and the dotted lines in Figure~2,
respectively. The obtained $\chi^2_{red}$ and the timescale for each
lightcurve and $EM$ were similar between the two
assumptions. Therefore both of the possibilities cannot be
discriminated. As for $EM$, both of the fits show no decay or very
long time decay, which is not seen in the usual solar flares. The
constant feature in $EM$ makes the quasi-static model
unacceptable. Since we cannot derive the aspect ratio of Flare II by
fitting with quasi-static cooling model, assuming the ratio derived in
Flare I and that radiative cooling is dominant, we deduced the plasma
parameters as shown in Table~3. Here, we derived the values assuming
the reheating scenario (the former assumption). The results show that
the plasma density and the volume remain roughly constant from Flare
I. This supports that Flare II resulted from reheating of the plasma
created by Flare I.


As for Flare III, because of the poor statistics and short observed
period, the fits to lightcurve with the exponential model give no
constrain between the above two possibilities. The results are shown
in Table~\ref{fit} and Figure~2. We derived the plasma parameters of
Flare III in the same way for Flare II, as shown in Table~3. These
values are similar to those in Flare I and II. All these are
consistent with the scenario that a quasi-periodic reheating makes the
triple flare. The heating interval is $\sim$20 hour. The loop size is
approximately constant through the three flares and is as large as
$\sim$14 $R_\odot$. The periodicity and the large-scale magnetic
structure support a scenario that an interaction between the star and
the disk occurred by the differential rotation and reheated the same
loop periodically (e.g., Hayashi, Shibata, \& Matsumoto 1996). The
detail will be presented in Paper II (Montmerle et al. 1999).


\small

\begin{deluxetable}{cccccc}
\tablecaption{Fitting of cooling phase of flares}
\tablehead{
\colhead{model expression } & \colhead{${\cal F}$} & 
\colhead{$\alpha$} & \colhead{${\cal F}_{\mathrm{p}}$}   & \colhead{$\tau$}
& \colhead{$\chi^2_{red}$}  \\  
 & & & & \colhead{[ks]} & \colhead{(d.o.f.)}\\
\colhead{[1]} & \colhead{[2]} & \colhead{[3]} & \colhead{[4]} & \colhead{[5]} 
& \colhead{[6]} }
\startdata
${\cal F}_{\mathrm{I,qs}}(t)={\cal F}_{\mathrm{p}}  \times  \{ 1+(t-t_{\mathrm{I}})/3\tau 
\} ^{-\alpha}$  &
$I$ [cnts\,ks$^{-1}$] 		& 4    & $172 \pm 5$   & $31 \pm 1$ & 0.8 (45)\\ 
& $kT$ [keV]     		& 8/7   & $5.6 \pm 0.6$ & [31]       & 0.5 (2)\\ 
& $EM$ [10$^{54}$\,cm$^{-3}$] 	& 26/7 & $10.4 \pm 0.9$& [31]       & 0.4 (2) \\
\tableline  
${\cal F}_{\mathrm{II}}(t)={\cal F}_{\mathrm{p}}  \times  exp \{ -(t-t_{\mathrm{II}})/\tau\}$  & $I$ 
[cnts\,ks$^{-1}$]
& --	& $62 \pm 3$    & $33 \pm 3$  & 1.7 (47)\\ 
& $kT$ [keV] 			& --	& $4 \pm 1$ & $50 \pm 20$ & 1.5 (1)\\  
& $EM$ [10$^{54}$\,cm$^{-3}$] 	& -- & $3 \pm 1$ & $>$ 53  & 0.4 (1) \\
${\cal F}_{\mathrm{I+II}}(t)={\cal F}_{\mathrm{I,qs}}(t) + {\cal F}_{\mathrm{II}}(t) $   & 
$I$ [cnts\,ks$^{-1}$]
	& --		& $46 \pm 3$ & $29 \pm 2$  & 1.6 (47)\\ 
& $EM$ [10$^{54}$\,cm$^{-3}$] 	& -- & $1.6 \pm 0.8$ & $>$ 54  & 0.6 (1) \\
\tableline  
${\cal F}_{\mathrm{III}}(t)={\cal F}_{\mathrm{p}}  \times  exp \{ -(t-t_{\mathrm{III}})/\tau\}$  & $I$ 
[cnts\,ks$^{-1}$]
& --	& $35 \pm 2$    & $60 \pm 20$  & 1.4 (24)\\ 
& $kT$ [keV] 			& --		& $5 \pm 3$ & 55 ($>$ 16)  & $\chi^2=0$ (0)\\ 
& $EM$ [10$^{54}$\,cm$^{-3}$] 	& --	& $2.5 \pm 0.7$ & 71 ($>$ 23)  & 
$\chi^2=0$ (0)\\
${\cal F}_{\mathrm{I+II+III}}(t)={\cal F}_{\mathrm{I+II}}(t)+{\cal F}_{\mathrm{III}}(t)$  & $I$ 
[cnts\,ks$^{-1}$]
& --	& $28 \pm 2$    & $60 \pm 20$  & 1.4 (24)\\ 
\enddata
\tablecomments{ 
Panel 1 gives the quasi-static fit for Flare I. Panel 2 (Panel 3)
gives the exponential fit for Flare II (Flare III): first assuming
that the flares are reheating of the same plasma volume as the
previous flare; second assuming that the three flares are independent
(the previous flares contribute to the latter
flares). $t_{\mathrm{I}}$, $t_{\mathrm{II}}$, and $t_{\mathrm{III}}$
are the beginning time of Flare I, II, and III, respectively. We
assume only positive values for the decay time (Column [5]). The
errors are at 90 \% confidence level. }
\label{fit} 
\end{deluxetable} 

\normalsize


\begin{deluxetable}{cccccccc}
\tablecaption{Physical parameters for the triple flare}
\small 
\tablehead{ 
\colhead{Flare}	& \colhead{$a$} 	& \colhead{$L$}		& \colhead{$n_e$}	& 
\colhead{$B$} & \colhead{$L_{X,peak}$}	& \colhead{$\tau$}  & 
\colhead{$E_{tot}$} \\
\colhead{}	& 		& \colhead{[R$_\odot$]}	& \colhead{[10$^{11}$\,cm$^{-3}$]} & 
\colhead{[G]}	&
\colhead{[10$^{31}$\,erg\,s$^{-1}$]}	& \colhead{[ks]}	& 
\colhead{[10$^{36}$\,erg]}}
\startdata 
I	& 0.07$\pm$0.02	& 14$\pm$2	& 0.5$\pm$0.1	& 150$\pm$20	& 20$\pm$1	& 31$\pm$1	& 6.0$\pm$0.5 \\
II	&[0.07$\pm$0.02]& 11$\pm$5	& 0.4$\pm$0.1	& 120$\pm$30	& 8$\pm$1 	& 33$\pm$3	& 2.8$\pm$0.6 \\
III	&[0.07$\pm$0.02]& 15$\pm$10	& 0.3$\pm$0.2	& 100$\pm$60	& 4.6$\pm$0.5 	& 60$\pm20$	& 2.7$\pm$1.0  \\
\enddata
\tablecomments{
$a$: diameter-to-length ratio of the loop; 
$L$: loop length; $n_e$: electron density; 
$B$: minimum value of the magnetic field confining the loop plasma;
$L_{X,peak}$: X-ray luminosity at the flare peak in the 0.1--100 keV band;
$\tau$: characteristic decay time of the lightcurve;
$E_{tot}$: total energy released during the cooling phase in X-ray wavelength.
}
\label{physical_parameters}
\end{deluxetable}
\normalsize

\subsection{Comparison with Other Flares}

Among YSOs, TTSs have been known for strong X-ray time variabilities
since the $Einstein$ Observatory discovered them (see \S1). At any
given moment, 5--10 \% are caught in a high-amplitude flare with
timescales of hours (Feigelson \& Montmerle 1999). The
most recent example is that of V773 Tau, which exhibits day-long
flares with $L_{X,peak} \sim$ 2--10 $\times 10^{32}$ erg s$^{-1}$ and
very high temperatures of $\sim 10^8$ K (Tsuboi et al. 1998). Other
examples of bright TTS X-ray flares are P1724, a WTTS in Orion
($L_{X,peak} \sim 2 \times 10^{33}$ erg s$^{-1}$; Preibisch,
Neuha\"user, \& Alcal\'a 1995), and LkH$\alpha$92, a CTTS in Taurus
(Preibisch, Zinnecker, \& Schmitt 1993). These X-ray properties 
resemble those of RS CVn systems.

Recently, a dozen protostars have been detected in X-rays, and four of
those showed evidence for flaring (RCrA core; Koyama et al. 1996,
EL29; Kamata et al. 1997, YLW15; Grosso et al. 1997, and SSV63 E+W;
Ozawa et al. 1999). In the ``superflare'' of YLW15 (Grosso et
al. 1997), an enormous X-ray luminosity was recorded during a few
hours. If we adopt the absorption we derived ($N_H =$ 3$\times10^{22}$
cm$^{-2}$), the absorption-corrected X-ray luminosity in 0.1--100 keV
band is $L_{X,peak} \sim 10^{34}$ erg\,s$^{-1}$. The ``triple flare''
we detected in this observation does not reach the same level as the
``superflare'': $L_{X,peak} = 5-20 \times 10^{31}$
erg s$^{-1}$ in the same X-ray band.

To compare the flare properties of our triple flare from YLW15 with
other flare sources, including RS CVns, we selected bright flare
sources as seen in Table 4. All the flare sources have a
well-determined temperature using a wide range of energy band of
$Tenma, Ginga$, and $ASCA$ satellites. Since the samples of
YSO flares were less, we added two TTS flares; the
flares on ROXs21 and SR24N detected in our observations (see \S2 and
\S3). We analyzed them using GIS data. The densities and volumes for
all the sources were derived assuming that radiative cooling is
dominant.


As a result, we found that although less energetic than the
``superflare'', with total energies in excess of 3--6 $\times 10^{36}$
ergs, the triple flare are on the high end of the energy distribution
for protostellar flares. While the plasma densities, temperatures,
and then derived equipartition magnetic fields are typical of stellar
X-ray flares, the emitting volume is huge; it exceeds those in
binary systems of RS CVns by a few orders of magnitude.





\small

\begin{deluxetable}{cccccccccc}
\tablecaption{Comparison with the other flares}
\tablehead{
\colhead{Source} & \colhead{SED} & \colhead{$n_e$} & \colhead{$V$}	& 
\colhead{$\tau$} 	& \colhead{$kT_{peak}$}	& \colhead{$EM_{peak}$}	& 
\colhead{$E_{tot}$} 	& \colhead{ref.}\\
       &	   & \colhead{[10$^{11}$\,cm$^{-3}$]} & \colhead{[10$^{33}\,cm^3$]} & 
\colhead{[ks]}   & \colhead{[keV]}  	& \colhead{[10$^{55}$\,cm$^{-3}$]} & 
\colhead{[10$^{36}$ erg]} & }
\startdata
YLW15(I) & I   & 0.5	& 3	& 31		& 6	& 0.7 & 6 & \\
YLW15(II) & I   & 0.4	& 2	& 33		& 5	& 0.3 & 3 & \\
YLW15(III) & I   & 0.3	& 2	& 60		& 4	& 0.2 & 3 & \\
EL29     & I    & 1	& 0.2	& 8	& 4		& 0.3 & 0.4 & (1)\\
RCrA     & I    & 1	& 0.04	& 20	& 6		& 0.04 & 0.2 & (2)\\	
SSV63E+W & I    & 1	& 2	& 10	& 6		& 2 & 4 & (3)\\	
\tableline 
SR24N   & II	& 5	& 0.1	& 6	& 7		& 3   & 6 & \\
\tableline 
ROXs31	& III	& 3	& 0.6	& 7	& 6		& 6   & 13 & \\	
ROX7    & III	& 3	& 0.01	& 6	& 3		& 0.1 & 0.1 & (1)\\
V773Tau & III	& 3	& 0.6	& 8	& 10		& 5   & 8 & (4)\\	
\tableline 
Algol 	&	& 1	& 0.09	& 20	& 5 		& 0.1 & 0.6 & (5)\\
$\pi$ Peg & 	& 8	& 0.006	& 2	& 6	 	& 0.3 & 0.2 & (6)\\
AU Mic   &	& 6	& 0.002	& 2	& 5		& 0.06	& 0.03 & (7)\\
\enddata
\tablecomments{
reference: (1) Kamata et al. 1997\\
(2) Koyama et al. 1996 and Hamaguchi private comm.\\
(3) Ozawa et al. 1999\\
(4) Tsuboi et al. 1998\\
(5) Stern et al. 1992\\
(6) Doyle et al. 1991\\
(7) Cully et al. 1993
}
\label{comparison} 
\end{deluxetable}


\normalsize

\section{Summary and Conclusions}

In the course of a long exposure of the $\rho$ Oph cloud with {\sl
ASCA}, we found evidence for a `triple flare' from the Class~I
protostar YLW15. This triple flare is the first example of its kind;
it shows an approximate periodicity of $\sim 20$ hours.  Each event
shows a clear decrease in the temperature, followed by reheating, with
$kT_{peak} \sim$ 4--6 keV, and luminosity $L_{X,peak} \sim$ 5--20 $\times
10^{31}$ erg s$^{-1}$. Apart from the periodicity, the characteristics
of the flares are among the brightest X-ray detections of Class~I
protostars.

A fitting with the quasi-static model (VM), which
is based on solar flares, reproduces the first flare well, and it suggests
that the plasma cools mainly radiatively, having semi-circular shape
with length $\sim 14\,R_\odot$ (radius of the circle $R \sim
4.5\,R_\odot$) and aspect ratio $\sim 0.07$. The minimum value of
the confining field is $B \sim 150$ G. The two subsequent flares have
weaker intensity than the first one but consistent with the reheating
of basically the same magnetic structure as the first flare. The
plasma volume is huge; a few orders of magnitude larger than the typical
flares in RS CVns. 

The fact that the X-ray flaring is periodic suggests that the cause
for the heating is periodic, hence is linked with rotation in the
inner parts of the protostar. The large size of the magnetic structure
and the periodicity support the scenario that the flaring episode has
originated in a star-disk interaction; differential rotation between
the star and disk would amplify and release the magnetic energy in one
rotation period or less, reheating the flare loop as observed in the
second and third flares.

\acknowledgements
The authors thank all the members of the $ASCA$ team whose efforts
made this observation and data analysis available. We also thank
Prof. Eric D. Feigelson, Mr. Kenji Hamaguchi, Dr. Mitsuru Hayashi,
Dr. Antonio Magazzu, Mr. Michael S. Sipior, and Prof. Kazunari Shibata
for many useful comments in the course of this work. Y.T.\
acknowledges the award of Research Fellowship of the Japan Society for
Young Scientists.


\newpage

\newpage
\appendix
\section*{APPENDIX\\ 
Determination of the Physical Parameters of the Flares}
\label{appendix}

\subsection*{1. Loop Parameters} 
	\label{VM}

We make use of the general treatment for solar-type flares, put
forward by van den Oord \& Mewe (1989, hereafter VM). The decrease in
the thermal energy of the cooling plasma is assumed to be caused by
radiative ($\tau_{\mathrm{r}}$) and conductive losses
($\tau_{\mathrm{c}}$): its decay time is thus
$1/\tau_{\mathrm{eff}}=1/\tau_{\mathrm{r}}+1/\tau_{\mathrm{c}}$. VM
assumed that the flare lightcurve and temperature decrease
exponentially with decay times $\tau_{\mathrm{d}}$ and
$\tau_{\mathrm{T}}$, respectively.  This effective cooling time is
related to observed time scales by
$1/\tau_{\mathrm{eff}}=7/8\tau_{\mathrm{T}}+1/2\tau_{\mathrm{d}}$.  We
assume here that the flare occurs in only one semi-circular loop with
constant section along the tube flux, radius $R$ ($R=L/\pi$; length
$L$), diameter-to-length ratio $a$, and volume $V$.  VM gives an
expression of $L$ versus $a$, depending on $\tau_{\mathrm{eff}}$, the
temperature, the emission measure (hereafter $EM$), and the ratio
$\tau_{\mathrm{r}}/\tau_{\mathrm{c}}$. Because of the assumed
exponential behavior of the lightcurve and temperature, the moment at
which this expression is applied is not important. The only
restriction is that both the temperature and the $EM$ started to
decrease.

Due to the low statistics, we have only time-sliced values of the 
temperature and the $EM$. Let us call $t_i$ ($t_f$) the beginning (end) of 
the time interval within which the temperature was 
estimated, we have: $\overline{T} = \int_{t_{i}}^{t_{f}} 
T(t^\prime)\,dt^\prime/(t_f-t_i)$. We use this relation to find the 
behavior of the temperature as a function of time. 

We now have a relation between $L$ and $a$, but the ratio
$\tau_{\mathrm{r}}/\tau_{\mathrm{c}}$ is unknown, and even worse may
change during the decay of the flare. An exception to this is when the
flare volume cools quasi-statically, evolving through a sequence of
{\it quasi-static equilibria}, where the loop satisfies scaling laws,
and where $\tau_{\mathrm{r}}/\tau_{\mathrm{c}}=cst$. Due to the
dependency of $\tau_{\mathrm{r}}$ and $\tau_{\mathrm{c}}$ on the
temperature and the $EM$, $\tau_{\mathrm{r}}/\tau_{\mathrm{c}}=cst$
implies $T^{13/4}/EM=cst$. VM gives in that case 
analytical expressions for several physical quantities versus time,
all depending on the quasi-static radiative time scale
($\tau_{\mathrm{r,qs}}$).  $\tau_{\mathrm{r,qs}}$ can be estimated
from the lightcurve which must be proportional to the radiative
loss. The temperature and the $EM$ can be fitted as described above
using this value of $\tau_{\mathrm{r,qs}}$, and give the peak values
$kT_{\mathrm{p,qs}}$, $EM_{\mathrm{p,qs}}$ (see Col.~[4]--[5] in our
Table \ref{fit}, panel~1 for details).

Using an expression for the radiative time (eq.[23a] of VM), the $EM$
(eq.[3] of VM) and the scaling law (SL[2], eq.[196] of VM), we
obtained the loop characteristics for the quasi-static model:
\begin{equation} 
a=1.38 \times (\tau_{\mathrm{r,qs}}/10\,ks)^{-1/2} \times  
(kT_{\mathrm{p,qs}}/keV)^{-33/16} \times
(EM_{\mathrm{p,qs}}/10^{54}\,cm^{-3})^{1/2}, 
\label{eq:a} 
\end{equation} 
\begin{equation} 
L=1.0\,R_\odot \times (\tau_{\mathrm{r,qs}}/10\,ks) \times 
(kT_{\mathrm{p,qs}}/keV)^{7/8},
\end{equation} 
\begin{equation} 
n_e=4.4 \times 10^{10}\,cm^{-3} \times (\tau_{\mathrm{r,qs}}/10\,ks)^{-1} 
\times
(kT_{\mathrm{p,qs}}/keV)^{3/4}.
\label{eq:n_e} 
\end{equation}

Using an expression for the ratio $\tau_{\mathrm{r}}/\tau_{\mathrm{c}}$ 
on page 252 of VM, we found 
$\tau_{\mathrm{r}}/\tau_{\mathrm{c}}=\mu \times f$, with the 
parameter $\mu$ depending only on the exponent of the temperature in the 
expression for the radiative loss (for temperature above 20\,MK, 
$\mu=0.18$), and the multiplicative function $f$ coming from the 
expression for the mean conductive energy loss (formula [7] of VM), 
which is equal to 4/7 for a loop with a constant section. Thus, 
$\tau_{\mathrm{r}}/\tau_{\mathrm{c}}=0.1$. \footnote{VM wrote 
$\tau_{\mathrm{r}}/\tau_{\mathrm{c}}=\mu=0.18$, and the analytical 
expression for the conductive energy loss in the quasi-static model 
without taking this multiplicative factor $4/7$ into account (see Table 
5 of VM).} In other words, the quasi-static model implies that 91\,$\%$ 
of the lost energy are radiative losses: radiation is the dominant 
energy loss process.

Assuming that the cooling is only radiative we used these 
simplified relations based on the exponential decay of the lightcurve, 
the temperature, and the $EM$:
\begin{equation}
n_e=4.4\times 10^{10}\,cm^{-3}\times (\tau_{\mathrm{d}}/10\,ks)^{-1} \times 
(kT_{\mathrm{p}}/keV)^{3/4},\mathrm{~for~kT>2\,keV},
\end{equation}
\begin{equation}
L=7.4\,R_\odot \times (a/0.07)^{-2/3}\times (\tau_{\mathrm{d}}/10\,ks)^{2/3} 
\times
(kT_{\mathrm{p}}/keV)^{-1/2}  \times (EM_{\mathrm{p}}/10^{54})^{1/3}, 
\end{equation}
with $kT_{\mathrm{p}}$ ($EM_{\mathrm{p}}$) the peak value of the temperature 
($EM$).

\subsection*{2. Magnetic Field}

Assuming equipartition between the magnetic pressure $B^2/8\pi$ 
and the ionized gas pressure $2n_e  kT$, we can obtain a minimum value of 
the magnetic field confining the emitting plasma using: 
\begin{equation}
B=28.4\,G \times 
(n_e/10^{10}\,cm^{-3})^{1/2} \times (kT/keV)^{1/2}
\label{eq:B}
\end{equation}

\subsection*{3. Released Energy}

For estimating the energy released by the flare during its cooling phase 
we need the peak luminosity value of this flare. 
As the lightcurve must be proportional to the intrinsic luminosity, 
we fit the time averaged intrinsic luminosities in the 0.1--100\,keV band 
(given in Table~1) with the same model used for the lightcurve fitting.
This gives the peak luminosity,  $L_{\mathrm{X,\,peak}}$, and a characteristic 
decay time $\tau$. 
Thus, the total energy released by this flare is:
\begin{equation}
 E_{\mathrm{tot}} \sim 10^{35}\,erg \times (L_{\mathrm{X,\,peak}}/10^{32}\,erg\,s^{-1}) \times
(\tau/ks)
\label{eq:E}
\end{equation}

\end{document}